\documentclass[10 pt]{article}
\usepackage[latin1]{inputenc}
\usepackage{amsmath}
\usepackage{amsfonts}
\usepackage{amssymb}
\usepackage{braket}
\usepackage{mathtools}
\usepackage{mathrsfs}
\usepackage{stmaryrd}
\usepackage[stable]{footmisc}
\usepackage{slashed}
\usepackage{calligra}
\usepackage{slashbox}
\usepackage{graphicx}
\usepackage[all]{xy}
\usepackage[amsmath,thmmarks]{ntheorem}
\usepackage{authblk}

\def\RR{{\mathbb R}}
\def\CC{{\mathbb C}}
\def\NN{{\mathbb N}}
\def\HH{{\mathbb H}}

\def\reel{{\operatorname{Re} \:}}

\def\bra{\langle}
\def\ket{\rangle}

\def\tr{\mbox{\rm Tr}}

\def\bea{\begin{eqnarray}}
\def\eea{\end{eqnarray}}

\def\be{\begin{equation}}
\def\ee{\end{equation}}
\def\og{{``}}
\def\fg{{''}}

\newtheorem{theorem}{Theorem}
\newtheorem{definition}{Definition}

\newtheorem{propo}{Proposition}
\newtheorem{cor}{Corollary}
\theoremstyle{nonumberplain}
\theorembodyfont{\normalfont}
\theoremseparator{:}
\theoremsymbol{$\P$}
\newtheorem{demo}{Proof}

\newcommand{\bigL}{\operatorname{Lex}}

\title{The disappearance of causality at small scale in almost-commutative manifolds}

\author[1]{Nadir Bizi \thanks{nadir.bizi@impmc.upmc.fr; nadir.bizi@ens.fr}}
\author[2]{Fabien Besnard \thanks{fabien.besnard@epf.fr}}
\affil[1]{\footnotesize Institut de minéralogie, de physique des matériaux et de cosmochimie, Université Pierre et Marie Curie, 4 place Jussieu, F-75252 Paris Cedex 05, France; École Normale Supérieure, International Center of Fundamental Physics, 24 Rue Lhomond, F-75005 Paris, France.}
\affil[2]{\footnotesize Pôle de recherche M.L. Paris, EPF, 3 bis rue Lakanal, F-92330 Sceaux, France.}

\begin{document}

\maketitle

\begin{abstract}
This paper continues the investigations of noncommutative ordered spaces put forward by one of the authors. These metaphoric spaces are defined dually by so-called \emph{isocones} which generalize to the noncommutative setting the convex cones of order-preserving functions. In this paper we will consider the case of isocones inside almost-commutative algebras of the form ${\cal C}(M)\otimes A_f$, with $M$ a compact metrizable space. We will  give a family of isocones in such an algebra with the property that every possible isocone is contained in exactly one member of the family. We conjecture that this family is in fact a complete classification, a hypothesis  related with the noncommutative Stone-Weierstrass conjecture.  We also obtain that every isocone in ${\cal C}(M)\otimes A_f$, with $A_f$ noncommutative, induces an order relation on $M$ with the property that every point in $M$ lies in a neighbourhood of incomparable points. Thus, if the causal order relation on spacetime is induced by an isocone in an almost-commutative (but not commutative) algebra, then causality must disappear at small scale.
\end{abstract}

\section{Introduction}
   
It is  probably unnecessary to recall the power of the beautiful noncommutative geometric interpretation of the standard model by Connes and Chamseddine (see \cite{VS} for an introduction and \cite{cc1} for some of the latest advances on the subject). This so-called \og spectral standard model\fg\ takes place on an almost-commutative algebra, that is a $C^*$-algebra of the form ${\cal C}(M)\otimes A_f$ where ${\cal C}(M)$ stands for the algebra of continuous functions on a compact manifold $M$, and $A_f$ is a finite-dimensional algebra, which is usually taken to be $\CC\oplus \HH\oplus M_3(\CC)$, but it is also possible to consider $\CC\oplus M_2(\CC)\oplus M_3(\CC)$ by slightly modifying the model (\cite{sz}), or even $M_2(\CC)\oplus M_4(\CC)$, in which case Pati-Salam unification is recovered (\cite{cc2}). However, in all these theories the metric, which is encoded in the Dirac operator, has an euclidean signature. This obstacle can be circumvented by the use of Wick rotations in the flat spacetime which is the arena for particle physics at today's accessible energies. However, this tool is not available in full generality. Hence a Lorentzian version of noncommutative geometry is called for, both from a physical and a mathematical perspective. While there already exist pioneering works in that direction (\cite{stro}, \cite{ps}, \cite{dpr}), a complete theory is still lacking. Moreover, most of these approaches aim at a definition of semi-Riemannian noncommutative geometry in general, though the Lorentzian case might be special.

In fact, we do know that it is special already in the commutative case, for the Lorentzian signature (that is $(1,n-1)$ or $(n-1,1)$) is the only one which allows (in general only locally) the definition of a partial order relation on the set of events. In other words, causality is a strictly Lorentzian phenomenon. Moreover, causality determines the conformal structure of spacetime, that is the metric up to a scalar field. It is therefore tempting to seek a Lorentzian formulation of noncommutative geometry which puts causality to the forefront. Two such programmes exist: the one of Franco (see \cite{fe} for a review), and the $I^*$-algebra approach first put forward in \cite{bes1}. Both share some common points but differ in one essential respect: Franco's  approach aims at defining causality    from the metric (encapsulated in the Dirac operator in noncommutative geometry) and some additional information, while in the theory of $I^*$-algebra our goal would be to reconstruct the metric from causality and the conformal factor (in a spirit quite similar to the causal set approach to quantum gravity, see \cite{sor}).

The theory of $I^*$-algebras is only in its infancy, with essentially only two questions dealt with to date. The first is the  definition of what  noncommutative topological ordered spaces should be. The tentative answer put forward in \cite{bes1} is that these corresponds by Gelfand duality to certain convex cones in $C^*$-algebras, called isocones. The couple $(I,A)$ where $I$ is an isocone in the $C^*$-algebra $A$ is called an $I^*$-algebra. We will briefly review $I^*$-algebras in section \ref{istar}. The second question already answered is the classification of all finite-dimensional $I^*$-algebras (\cite{bes2}).

In this paper we go one step further, with the investigation of the almost-commutative case, the importance of which has been recalled above. In section \ref{acisoc} we will consider almost-commutative algebras of the form ${\cal C}(M)\otimes A_f$ where $M$ is a compact metrizable space. We will exhibit for each such algebra a family of isocones such that every isocone in ${\cal C}(M)\otimes A_f$ is contained in one and only one member of the family. In fact, we will explain why we expect that this family is exhaustive, constituting a complete classification of almost-commutative $I^*$-algebras. We will prove in passing a very striking result: if the algebra $A_f$ is not commutative, and if $\preceq$ is a closed partial order  on $M$ (we will explain why this closure condition is  necessary), there exists an isocone in ${\cal C}(M)\otimes A_f$ which induces the order $\preceq$ on $M$ if and only if the strict order relation $\prec$ is also closed. This surprising condition means that for every point $p$ in the manifold, there exists a neighbourhood of $p$ in which no other point is comparable to $p$ for $\preceq$. More emphatically: causality must disappear at small scale. We will end the paper with a discussion of this result, and in particular of its agreement with the findings of \cite{lizzi} and \cite{asz} in spite of completely different methods.

\section{Noncommutative ordered spaces}\label{istar}
\subsection{Topological ordered spaces}
Let us recall a few definitions and notations. A \emph{topological ordered set} $M$ is a set which is at the same time a topological space and a partially ordered set (\emph{poset} for short). A map $f : M\rightarrow N$ between two posets is said to be an \emph{isotony} when it preserves the order, that is

\be
x\preceq_M y\Rightarrow f(x)\preceq_Nf(y)
\ee

for all $x,y\in M$, and writing $\preceq_M,\preceq_N$ for the partial order relations on the respective posets. A map is said to be \emph{strictly increasing} if

\be
x\prec_My\Rightarrow f(x)\prec_Nf(y)
\ee

where the strict partial order symbol is defined such that $a\prec b\Leftrightarrow a\preceq b$ and $a\not=b$.

We will be interested in continuous isotonies from $M$ to $\RR$. The set of such maps will be written $I(M)$, or $I(M,\preceq)$ if we want to emphasize the ordering we consider. It is clear that Gelfand theory cannot work unless we have enough information in $I(M)$ to recover $\preceq$. We are thus led to assume that

\be
(\forall f\in I(M),\ f(x)\le f(y))\Rightarrow x\preceq y\label{deftoposet}
\ee

A topological ordered set satisfying (\ref{deftoposet}) is said to be \emph{completely separated}. In this paper, following \cite{bes1}, a completely separated topological ordered space will be called a \emph{toposet} for the sake of brevity.

As we have said, the toposet property is a necessary condition to expect a reconstruction theorem. It is hence desirable to have some examples of this important kind of spaces. First, it is a well-known property of Minkowski spacetime that if two events $p$ and $q$ are not causally related, then there exist two observers $O_1$ and $O_2$ such that $p$ will chronologically precede $q$ for $O_1$ and the other way around for $O_2$. This immediately entails that the Minkowski spacetime is a toposet for the causal order relation. In fact we can be much more general. A spacetime  is causally simple if 1) it does not contain closed causal curves, and 2) the causal future and past of every point is closed (for the original definition see \cite{kp}, for this simplified form see \cite{bs}). One can prove (see \cite{mingstrK1} th. 2.2 or \cite{minguzzisanchez} lemma 3.67) that the second point is equivalent to $\preceq$ being closed in $M\times M$. Hence a spacetime is causally simple if and only if the causal relation $\preceq$ is a closed partial order. This is the second strongest causality condition after global hyperbolicity. It turns out that it is equivalent to the toposet property. To see this we can use Levin's theorem (\cite{levin}) which we now recall.

\begin{theorem} {\bf (Levin's theorem)} Let $M$ be a second countable locally compact Haussdorf space and $\preceq$ a closed partial order on $M$. Then there exists a real continuous strictly increasing function on $M$. Moreover, if $f(x)\le f(y)$ for every such functions, then $x\preceq y$.
\end{theorem}

A few remarks are in order. First, the second part immediately entails that $M$ is a toposet. This part is not in the original formulation of Levin but is an easy consequence of his proof, as remarked in \cite{eo}. Moreover, this theorem still holds with the partial order replaced with a mere pre-order, up to an appropriate redefinition of strictly increasing functions. It is remarkable that this theorem arose in the context of mathematical economics (where strictly increasing functions are known as utilities) and has only recently been put to use in causality theory (see \cite{minguzzi}). Finally let us remark that if $M$ is compact, the case in which we will work from the next section on, the hypotheses \og Haussdorf\fg\  and \og second-countable\fg\ are equivalent to metrizability by Urysohn's metrization theorem. 


Let us mention here an immediate corollary which will be useful to us in the sequel.

\begin{cor}\label{Levin} Let $(M,\preceq)$ be a compact metrizable ordered space such that $\prec$ is closed in $M\times M$. Then $(M,\preceq)$ is a toposet and there exists a continuous isotony on $M$ such that
\be
x\prec y\Rightarrow g(y)-g(x) \geq 1
\ee
for all $x,y\in M$.
\end{cor}
\begin{demo} First $\preceq$ is obviously closed since it is the union of $\prec$ and the diagonal, two closed subsets of $M\times M$. Hence by Levin's theorem $M$ is a toposet and there exists a strictly increasing function $v$ on $M$. If $\prec$ is the empty relation, the corollary is trivial. If not, then $m=\inf_{x\prec y}(v(y)-v(x))$ is finite by compactness and $g=v/m$ is the function we seek.
\end{demo}

\subsection{Gelfand theory for toposets}
We now recall the following definition from \cite{bes3}:

\begin{definition}\label{defisoc} Let $A$ be a $C^*$-algebra with unit $1$.  A subset $I$ of $\reel (A)$ which satisfies :
\begin{enumerate}
\item $\forall x\in\RR$, $x.1\in I$.\label{constants}
\item $\forall a,a'\in I$, $a+a'\in I$,\label{somme} 
\item $\forall a\in I$, $\forall f\in I(\RR)$, $f(a)\in I$,\label{isotcalc}
\item $\overline{I}=I$ ($I$ is norm-closed),\label{ferme}
\end{enumerate}

will be called a pre-isocone. A pre-isocone will be called an isocone if it moreover satisfies 
\begin{enumerate}\setcounter{enumi}{5}
\item\label{dense} $\overline{I-I}=\reel(A)$
\end{enumerate}

A couple $(I,A)$ when $I$ is an isocone of $A$ is called an $I^*$-algebra. The set of isocones of $A$ is denoted $\mathcal{I}(A)$.
\end{definition}

An equivalent definition can be found in \cite{bes1} under the name \og weak $I^*$-algebra\fg. The equivalence between the two definitions is proved in \cite{bes3}. Given an isocone $I$, we can define a partial order structure on $S(A)$, the state space of $A$, in the following way : given two  states $\phi,\psi$, one has by definition 

\be
\phi\le_I\psi\Longleftrightarrow \forall a\in I,\ \phi(a)\le\psi(a)\label{deforder}
\ee

Note that this would define a pre-order given \emph{any} subset $I$ of $\reel A$, but the condition \ref{dense} ensures that it is a partial order. Also, since the evaluation $ev_a$ at $a$ is continuous, (\ref{deforder}) defines a toposet structure on $S(A)$. We now restrict this ordering to the pure state space $P(A)$ of $A$, which is compact for the weak $*$-topology. We have the following theorem (\cite{bes1}), which we quote here informally:

\begin{theorem} {\bf (Gelfand duality for commutative $I^*$-algebras)} Commutative $I^*$-algebras are exactly those of the form $(I(M),{\cal C}(M))$, with $M$ a compact toposet.
\end{theorem}
We note that with the correct notion of morphisms, the above theorem can be raised to the level of a dual equivalence of categories, exactly like the original  Gelfand-Naimark theorem for commutative $C^*$-algebras, of which it is a generalization. For details on this point, see \cite{bes1}.

Naturally, the toposet $M$ is recovered from the commutative algebra by taking the pure state space (which is equal to the character space in this case), the ordering on $M$ is given by (\ref{deforder}), and the isocone exactly corresponds to $I(M,\le_I)$ via the Gelfand transform $a\mapsto ev_a$. This last point is the only one which is not entirely obvious. It follows from an appropriate variation of the Stone-Weïerstrass theorem (\cite{bes2}).

Remark: Some might worry that the compactness hypothesis is too strong since compact spacetimes cannot be causal. This is not true as such: spacetimes \emph{without boundary} cannot be causal. The Gelfand duality of this section and the results about almost-commutative manifolds in section 3 will thus apply either to a causal compactification of the entire spacetime, such as Penrose compactification, or, in greater generality, to a compact neighbourhood of a given point in some spacetime. Of course we will have boundaries in both cases.

\subsection{Noncommutative $I^*$-algebras}

The theorem of the previous section provides a serious mathematical motivation for considering noncommutative $I^*$-algebras as the correct dual objects of the metaphoric noncommutative ordered spaces, with noncommutative spacetimes among them.

On the physical side, we can also give some arguments for taking that road. First, we can justify taking the pure state space $P(A)$ of the algebra of observables as a substitute for spacetime (instead of the character space, the primitive spectrum or any other space attached to $A$). Indeed, in a commutative  spacetime $M$, we can (redundantly) characterize events by the values taken on them by every possible causal function $f\in I(M)$. In the noncommutative case there is no way to simultaneously assign a definite value to the measurement of all \og quantum\fg\ causal observables. In other words there are no events anymore. However the next best thing is a pure state, since at least one causal observable has a definite value in it. Now we ask for a partial ordering on $P(A)$ as a replacement for the causal order on $M$. This might seem a little naive at first. However, this requirement is much milder than one might think, since  \emph{any} set of observables induces a pre-order on $P(A)$, as we already remarked. Hence we are just asking that this set is large enough to separate the pure states. 

Now we make a saturation requirement: we take as the set of causal observables the set containing all observables which induce an isotone evaluation map on the pure states. This might sound like a vicious circle, but in fact it is just a biduality property: we want the set $I\subset \reel A$ to satisfy 

\begin{enumerate}
\item $I$ induces a closed partial order $\le_I$ on $P(A)$, and
\item $I=\{a\in \reel A| ev_a$ is isotone for $\le_I\}$.
\end{enumerate}

In the commutative case, isocones are precisely the sets of real functions which satisfy this biduality property, and as we remarked at the end of the previous section, this is proved thanks to some variation of the Stone-Weïerstrass theorem. Now we conjecture that this property is true in the noncommutative case as well.
\smallbreak

{\bf Saturation conjecture for $I^*$-algebra :} Let $(I,A)$ be an $I^*$-algebra. Then  $I=\{a\in \reel A| ev_a$ is isotone on $P(A)$ for $\le_I\}$.

\smallbreak

This conjecture is of course true in the commutative case, and it can be shown to hold also in the finite-dimensional case. We note also that if we replace the pure state space by the state space $S(A)$, then the conjecture is true by an immediate application of the Hahn-Banach theorem. We believe that the saturation conjecture is closely tied with the noncommutative Stone-Weïerstrass conjecture for Jordan algebras \cite{ncswJ}, and is likely to be equivalent to it given the results in \cite{brown}.

\subsection{Finite-dimensional $I^*$-algebras}

It is possible to classify finite-dimensional $I^*$-algebras (\cite{bes3}). Before recalling the classification theorem we will need to introduce some terminology.

\begin{definition}\label{lexicosum} Let $(P,\preceq)$ be a \emph{finite} poset and for each $x\in P$ let $(I_x,A_x)$ be an $I^*$-algebra. We set $I=\bigoplus_{x\in P}I_x$, $A=\bigoplus_{x\in P}A_x$, and we write elements of $A$ in the form $(a_x)_{x\in P}$. We define 
$$\bigL_{x\in P}I_x=\{a\in I|\forall x,y\in P, x\prec y\Rightarrow \max\sigma(a_x)\le \min\sigma(a_y)\}$$
which we call the lexicographic sum of $(I_x)_{x\in P}$.
\end{definition}

We will generalize this definition in the next section, replacing the finiteness condition on $P$ with some topological assumptions. We will then show that $\bigL_{x\in P}I_x$ is an isocone of $A$.

The name comes from the fact that the lexicographic sum $L=\bigL_{x\in P}I_x$ induces the lexicographic order on the pure state space $P(A)\simeq \coprod_{x\in P}P(A_x)$. That is, if we write $(x,\phi)$ for an element of the piece labeled $x$ in the disjoint sum, one has

\be
(x,\phi)\le_L (y,\psi)\Longleftrightarrow x\prec y\mbox{ or }(x=y\mbox{ and }\phi\le_{I_x}\psi)
\ee

We can now state the classification result. Recall that every finite-dimensional $C^*$-algebra is isomorphic to a direct sum of matrix algebras.

\begin{theorem}
Let $I$ be an isocone in the finite-dimensional $C^*$-algebra $A=\bigoplus_{x\in P}M_{n_x}(\CC)$, with $P=\{1;\ldots;k\}$, $k\in\NN^*$, $n_x\in\NN^*$. Then there exists a poset structure on $P$ such that $I=\bigL_{x\in P} I_x$. Moreover, if $n_x\not=2$ then $I_x=\reel M_{n_x}(\CC)$, and if $n_x=2$ then $I_x$ is any closed convex cone of $\reel M_2(\CC)$ which contains the constants and has a nonempty interior.
\end{theorem}

We refer to \cite{bes3} for the proof. Let us take a closer look at the case $n_x=2$ which is special. In any matrix algebra the pure states are of the form $a\mapsto \tr(ap_\xi)=\bra \xi,a\xi\ket$ where $p_\xi$ is the rank one  projection on the line generated by the normalized vector $\xi$. Hence $P(M_n(\CC))\simeq \CC P^{n-1}$. In the case $n_x=2$ the rank one projections form the two-sphere $S=\{p\in \reel M_2(\CC)|\tr(p)=\tr(p^2)=1\}$ of the hyperplane of trace one hermitian matrices, which is centered on $I_2/2$ and has Frobenius norm $\sqrt{2}/2$. An isocone $I$ in $M_2(\CC)$ is then entirely characterized by its intersection $K=I\cap S$ with $S$, which is a closed and geodesically convex subset of $S$ with non-empty interior (it can easily be seen to be either the whole $S$ or a subset of a closed hemisphere). The order $\le_I$ on $P(M_2(\CC))\simeq S$ can then be interpreted in the following way: for any $p,q\in S$

\be
p\le_I q\Longleftrightarrow \forall x\in K,\ d(x,p)\ge d(x,q)
\ee

where $d$ is the geodesic distance on $S$.

\begin{figure}[hbtp]
\begin{center}
\includegraphics[scale=0.5]{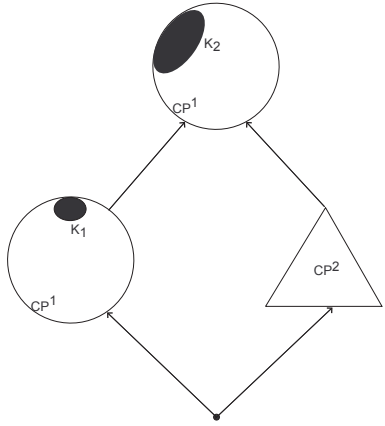}
\caption{In this example we take the algebra $A=\CC\oplus M_2(\CC)\oplus M_3(\CC)\oplus M_2(\CC)$. The poset $P=\{1;2;3;4\}$ is ordered according to a $1\preceq  2,3\preceq  4$ with $2\parallel 3$. The pure state space is $\{ev_1\}\coprod \CC P^1\coprod \CC P^2\coprod \CC P^1$. The first $\CC P^1$ is ordered according to a geodesically convex subset $K_1$, the second one according to $K_2$. The point and the $\CC P^2$ are trivially ordered.}\label{fig1}
\end{center}
\end{figure}

An example of a more general isocone in a finite-dimensional algebra is given in figure \ref{fig1}.

Remark: For application of noncommutative geometry to particle physics it can be important to consider real $C^*$-algebras, like $\CC\oplus\HH\oplus M_3(\CC)$. It is easy to see that in the real case, $M_2(\RR)$ and $M_2(\HH)$ are exceptional along with $M_2(\CC)$. This is the only significant change in the classification of finite-dimensional $I^*$-algebras. This will be discussed elsewhere.


\section{Isocones in almost-commutative algebras}\label{acisoc}

Using the classification theorem in the finite dimensional case we will, in this section, classify all isocone-induced orders for almost-commutative algebras. But contrary to what is done for the finite-dimensional case, we will not classify the isocones \emph{per se}, for we do not have (yet) any uniqueness theorem relating isocone-induced orders to the isocones.

\subsection{Almost-commutative geometries}

We start this section with a few definitions and notations. 

An almost commutative algebra is the product of a commutative algebra and a finite-dimensional noncommutative algebra. It is thus of the form: 

\begin{equation*}
A=\mathcal{C}(M) \otimes A_{f} \cong \mathcal{C}(M,A_{f}),
\end{equation*}
with $M$ compact Hausdorff, and $A_{f}=\bigoplus_{k=1}^{K} M_{n_{k}}(\mathbb{C})$. Let $\pi_{k}$ denote the projection on the $k$-th component of the algebra. Let $F \in A$. The element $F$ can be written in terms of components as $F=(F_{k})_{k=1..K}$, with $F_{k} : x \longmapsto \pi_{k}(F(x)) \in \mathcal{C}(M,M_{n_{k}}(\mathbb{C}))$ in the $k$-th component of the algebra. The pure state space of $A$ is:

\begin{equation*}
P(A) \cong M \times P(A_{f}) = M \times \coprod_{k=1}^{K} \CC P^{n_{k}-1} \cong \coprod_{k=1}^{K} (M \times  \CC P^{n_{k}-1}),
\end{equation*}
where the unions are disjoint. We choose to see it as a bundle\footnote{A bundle is analogous to a fiber bundle, but such that the fiber space is not necessarily the same everywhere.} over $M_{K}= \llbracket 1,K \rrbracket \times M$. The elements of $P(A)$ will be denoted in the form: $(k,x,\xi) \in P(A)$, with $(k,x) \in M_{K}$, and $\xi \in \CC P^{n_{k}-1}$; their action on $A$ is given by:

\begin{equation*}
(k,x,\xi)(F) \equiv \xi^{\dagger}F_{k}(x)\xi,
\end{equation*}
where $\xi$ was obviously replaced by a unit vector representing it. 
For $I \in \mathcal{I}(A)$ and $(k,x,\xi),(l,y,\eta) \in P(A)$, we have:

\begin{equation*}
(k,x,\xi) \leq_{I} (l,y,\eta) \Leftrightarrow \forall F \in I: \xi^{\dagger}F_{k}(x)\xi \leq \eta^{\dagger}F_{l}(y)\eta.
\end{equation*}

We embed $\mathcal{C}(M_{K})$ in $A$ in the following way:

\begin{equation*}
\begin{split}
\mathcal{C}(M_{K}) & \longrightarrow A \\
f & \longmapsto \bigoplus_{k=1}^{K} f(k, \cdot) 1_{n_{k}};
\end{split}
\end{equation*}
the $k$-th component of $f$ is thus $f_{k} \equiv f(k, \cdot)$.

Finally, we define the evaluation map $ev_{x}$ at $x \in M$: 

\begin{equation*}
\begin{split}
ev_{x} : A & \longrightarrow A_{f} \\
F & \longmapsto F(x),
\end{split}
\end{equation*}
and we denote, for any subset $S$ of $A$: 

\begin{equation*}
S(x) \equiv ev_{x}(S)=\{F(x) | F \in S\}.
\end{equation*}
More generally, for a finite subset $P$ of $M$, we define the restriction mapping $ev_{P}$:

\begin{equation*}
\begin{split}
ev_{P} : A & \longrightarrow \bigoplus_{x \in P} (A_{f})_{x} \\
F & \longmapsto \bigoplus_{x \in P} F(x),
\end{split}
\end{equation*}
where the index $x$ on $(A_{f})_{x}$ is to indicate that it is the copy of $A_{f}$ associated to $x$. It is clear that $ev_{P}$ is a continuous $*$-morphism.

In the simple case where $K=1$, we have $A=\mathcal{C}(M) \otimes M_{n}(\mathbb{C})$. The pure state space reduces to $P(A) \cong M \times \CC P^{n-1}$, with its elements denoted in the form $(x,\xi)$. For $I \in \mathcal{I}(A)$ and $(x,\xi),(y,\eta) \in P(A)$, we have:

\begin{equation*}
(x,\xi) \leq_{I} (y,\eta) \Leftrightarrow \forall F \in I: \xi^{\dagger}F(x)\xi \leq \eta^{\dagger}F(y)\eta.
\end{equation*} 

Let us now classify such orders. This classification is done in two steps: we first prove that all orders on almost-commutative manifolds are lexicographic. We then determine the necessary and sufficient conditions on an arbitrary lexicographic order to be induced by an isocone. We will present this classification for the case $K=1$. Thus, throughout this section, $A$ always denotes the algebra $A=\mathcal{C}(M) \otimes M_{n}(\mathbb{C})$, with $n \geq 2$ and $M$ compact Hausdorff. The analogous results for a more general almost-commutative algebra are then stated, but the proofs are delayed to the appendix.

\subsection{Isocone-induced orders are lexicographic}

\begin{theorem}\label{Class MnC 1}  Let $I \in \mathcal{I}(A)$. Then there exists an order $\preceq_{M}$ on $M$ such that for all $(x,\xi),(y,\eta) \in P(A)$:
\begin{equation}
(x,\xi) \leq_{I} (y,\eta) \Leftrightarrow 
\begin{cases}
x \prec_{M} y \\
\text{or} \\
x=y \text{ and } \xi \leq_{I_{x}} \eta,
\end{cases}
\label{lexico1}
\end{equation}
where $I_{x}=\overline{I(x)} \in \mathcal{I}_{n}$.
\end{theorem}

Here $\mathcal{I}_{n}:=\mathcal{I}(M_{n}(\CC))$ denotes the set of isocones of $M_{n}(\CC)$

\begin{demo}
The key idea of the proof is to restrict the algebra to a finite subset of $M$, in order to obtain a finite-dimensional algebra, of which the isocones are all known.

Let $P$ be a finite subset of $M$, and $ev_{P}(A)=\bigoplus_{x \in P} M_{n}(\mathbb{C})_{x}$ the restriction of $A$ to $P$. Its pure state space is $P(ev_{P}(A)) \cong P \times \CC P^{n-1} \subset P(A)$. 

Let $I \in \mathcal{I}(A)$. One can prove easily, or infer from Theorem 9 in \cite{bes3} and the fact that $ev_{P}$ is a surjective $*$-morphism, that $I_{P}=\overline{ev_{P}(I)}$ is an isocone of $ev_{P}(A)$. Let $(x,\xi),(y,\eta) \in P \times \CC P^{n-1}$. We have:

\begin{equation*}
\begin{split}
(x,\xi) \leq_{I} (y,\eta) & \Leftrightarrow \forall F \in I: \; \xi^{\dagger}F(x)\xi \leq \eta^{\dagger}F(y)\eta \\
& \Leftrightarrow \forall F \in ev_{P}(I): \; \xi^{\dagger}F(x)\xi \leq \eta^{\dagger}F(y)\eta \\
& \Leftrightarrow \forall F \in I_{P}: \; \xi^{\dagger}F(x)\xi \leq \eta^{\dagger}F(y)\eta \\
& \Leftrightarrow (x,\xi) \leq_{I_{P}} (y,\eta);
\end{split}
\end{equation*}
going from the first to the second line is possible because $I$ and $I_{P}$ take the same values on $P$, going from the second to the third comes from the fact that the natural order on $\mathbb{R}$ is closed. Thus the orders $\leq_{I}$ and $\leq_{I_{P}}$ coincide on $P \times \CC P^{n-1}$.

We know that $I_{P}$ is a finite-dimensional isocone. Therefore, there exists an order $\preceq_{P}$ on $P$ such that:

\begin{equation}
(x,\xi) \leq_{I} (y,\eta) \Leftrightarrow 
\begin{cases}
x \prec_{P} y \\
\text{or} \\
x=y \text{ and } \xi \leq_{I_{x}} \eta,
\end{cases}
\label{lexico2}
\end{equation}
where $I_{x}=\pi_{x}(I_{P})$. 

We have:
\begin{equation*}
I_{x} = \pi_{x}(\overline{ev_{P}(I)}) \subset \overline{\pi_{x}(ev_{P}(I))} = \overline{I(x)}.
\end{equation*}
Conversely:
\begin{equation*}
I(x)  = \pi_{x}(ev_{P}(I)) \subset \pi_{x}(\overline{ev_{P}(I)}) = I_{x}.
\end{equation*}
Hence: $ I_{x}=\overline{I(x)}.$

Let $\xi_{0}$ be an arbitrary fixed element of $\CC P^{n-1}$. Consider the binary relation $\preceq_{M}$ on $M$ defined by:

\begin{equation}
\forall x,y \in M: \; x \preceq_{M} y \Leftrightarrow (x,\xi_{0}) \leq_{I} (y,\xi_{0}).
\label{M_order}
\end{equation}
This clearly defines a closed order on $M$. It can be readily seen from \eqref{lexico2} that for $x,y \in P$, we have $(x,\xi_{0}) \leq_{I} (y,\xi_{0}) \Leftrightarrow x \preceq_{P} y$. Therefore, the orders $\preceq_{M}$ and $\preceq_{P}$ coincide on $P$. The equivalence \eqref{lexico2} can now be rewritten:

\begin{equation}
\forall (x,\xi),(y,\eta) \in P \times \CC P^{n-1}: \;
(x,\xi) \leq_{I} (y,\eta) \Leftrightarrow 
\begin{cases}
x \prec_{M} y \\
\text{or} \\
x=y \text{ and } \xi \leq_{I_{x}} \eta,
\end{cases}
\end{equation}
This is true for all finite subsets $P$ of $M$, and the right-hand side is now independent of the choice of the finite subset. It is thus true for all points in the pure state space.
\end{demo}

\begin{figure}
\includegraphics[width=\columnwidth]{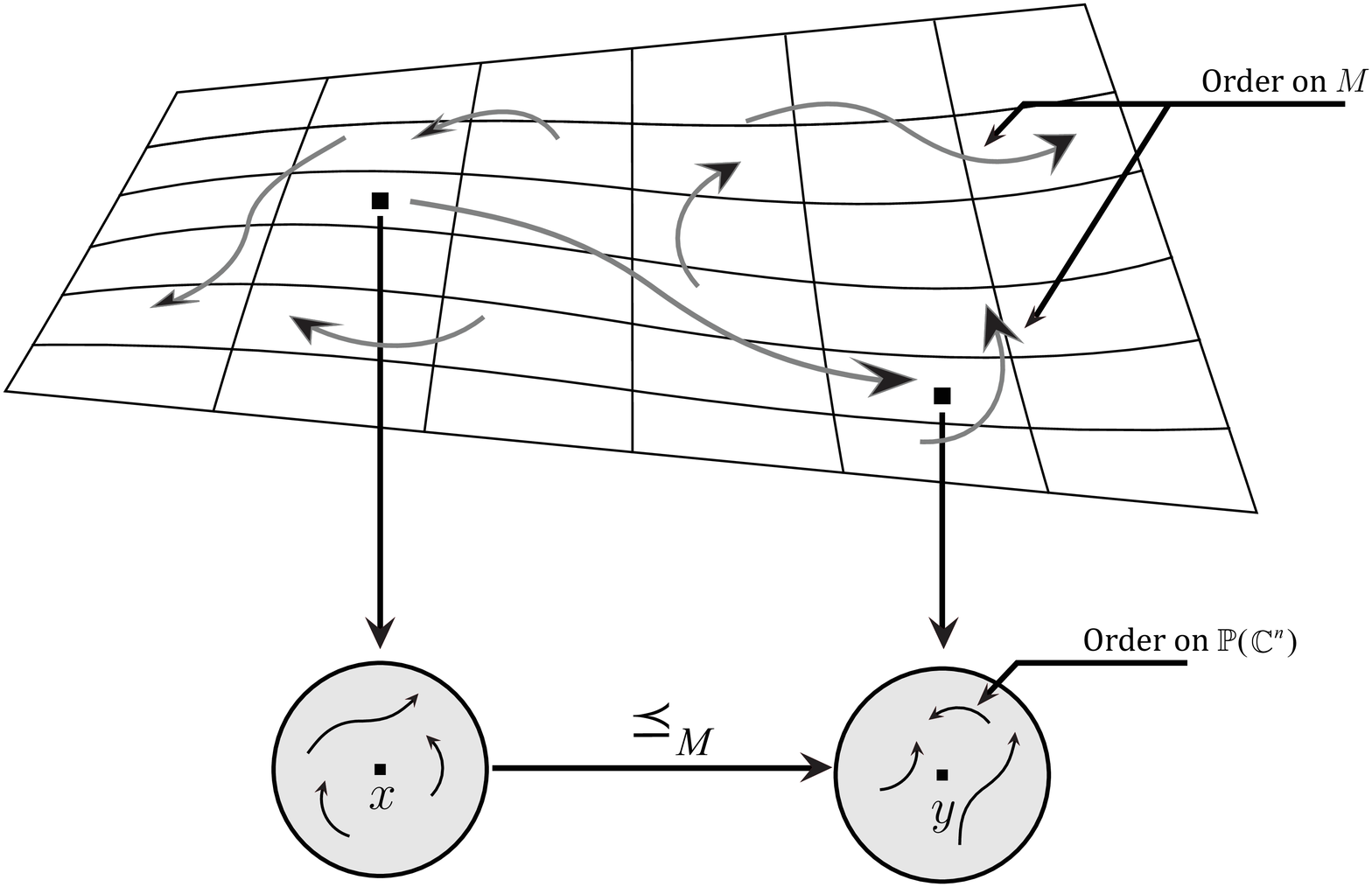}
\caption{The pure states are ordered on $M$, then on the internal space of each point, just as in the finite case.}
\label{LexicoOrder}
\end{figure}

Let us make some observations about the order $\preceq_{M}$. As can be seen from \eqref{M_order}, $\preceq_{M}$ is a closed order, and therefore $(M,\preceq_{M})$ is a compact toposet. Notice also that $\preceq_{M}$ is ultimately independent of $\xi_{0}$, due to the lexicographic nature of the order $\leq_{I}$.

Moreover, the strict order $\prec_{M}$ can be characterized in the following way:

\begin{equation}
\forall x,y \in M: \; x \prec_{M} y \Leftrightarrow \forall F \in I: \; \max \sigma(F(x)) \leq \min \sigma(F(y)).
\label{M_strictorder}
\end{equation}
Indeed, let $x,y \in M$. It is obvious that if $x \prec_{M} y$, then $(x,\xi) \leq_{I} (y,\eta)$ for all $\xi,\eta \in \CC P^{n-1}$. Conversely, suppose that $\forall \xi,\eta \in \CC P^{n-1}: \; (x,\xi) \leq_{I} (y,\eta)$. We have $x \preceq_{M} y$ thanks to \eqref{lexico1}. If $x=y$, then $\forall \xi,\eta \in \CC P^{n-1}: \; \xi \leq_{I_{x}} \eta$, and thus $\forall \xi,\eta \in \CC P^{n-1}: \; \xi = \eta$, which is absurd, given that $n \geq 2$. Therefore $x \prec_{M} y$. Thus:

\begin{equation*}
\begin{split}
x \prec_{M} y & \Leftrightarrow \forall \xi,\eta \in \CC P^{n-1}: \; (x,\xi) \leq_{I} (y,\eta) \\
& \Leftrightarrow \forall \xi,\eta \in \CC P^{n-1}, \; \forall F \in I: \; \xi^{\dagger}F(x)\xi \leq \eta^{\dagger}F(y)\eta \\
& \Leftrightarrow \forall F \in I: \; \max \sigma(F(x)) \leq \min \sigma(F(y)),
\end{split} 
\end{equation*}
hence \eqref{M_strictorder}. The most striking consequence of this characterization is that \emph{the strict order $\prec_{M}$ is closed}. This itself has important consequences, as we will see in section (4).

Let us give a simple example of such an order. Consider the space $M=[0,1]$, and $\lambda : [0,1] \rightarrow \mathbb{R}_{+}^{*} $ a positive function. The order $\preceq$ defined by:

\begin{equation*}
\forall x,y \in [0,1]: \; x \preceq y \Leftrightarrow \begin{cases}
y-x \geq \lambda(x) \\
\text{or} \\
x=y
\end{cases}
\end{equation*}
is such that the strict order $\prec$ is closed if and only if $\lambda$ is lower semi-continuous.

\begin{figure}[h!]
\centering
\includegraphics[width=0.5\columnwidth]{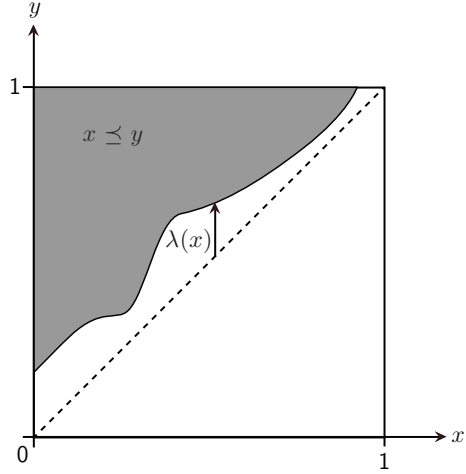}
\caption{The grayed out area is the graph of the strict order relation $\prec$}
\label{OrderOn0_1}
\end{figure}

We see that all isocone-induced orders are lexicographic, and are characterized by an order $\preceq_{M}$ on $M$, and a map of ``local'' isocones
\begin{equation*}
\begin{split}
L : M & \longrightarrow \mathcal{I}_{n} \\
x & \longmapsto I_{x}.
\end{split}
\end{equation*}  
We have to find the necessary and sufficient conditions on $\preceq_{M}$ and $L$ for their lexicographic combination to be isocone-induced. As it turns out, we already have half of these conditions: the condition on the order on $M$ is that $\prec_{M}$ be closed. We still need the conditions on the local isocone map. To state these conditions, we need to open a short parenthesis on multi-valued functions (also referred to as carriers), and state an important result from E. Michael.

\subsection{Multi-valued functions and the Michael selection theorem}

Let us first define multi-valued functions and selections. These definitions can be found in \cite{Michael}.

\begin{definition}
\leavevmode
\begin{itemize}
\item Let $X$ and $Y$ be topological spaces. A \emph{multi-valued function} from $X$ into $Y$ is a mapping from $X$ into the set $2^{Y}$ of non-empty subsets of $Y$.
\item Let $\varphi : X \longrightarrow 2^{Y}$ be a multi-valued function. A \emph{selection} for $\varphi$ is a continuous mapping $f \in \mathcal{C}(X,Y)$ such that $f(x) \in \varphi(x)$ for all $x \in X$.
\end{itemize}
\end{definition}
The set of selections of a given multi-valued function $\varphi$ is denoted $\mathscr{S}(\varphi)$.

Notions of continuity can be defined for multi-valued functions. In particular:

\begin{definition}
Let $\varphi : X \longrightarrow 2^{Y}$ be a multi-valued function. $\varphi$ is said to be \emph{lower hemi-continuous} (l.h.c.) if and only if the set $\{x \in X | \varphi(x) \cap V \ne \varnothing \}$ is open for all open $V \subset Y$.

Equivalently, $\varphi$ is l.h.c. if and only if $\forall x \in X, \forall y \in \varphi(x)$ and all $V \subset Y$ neighborhood of $y$, there exists a neighborhood $U$ of $x$ such that $\forall x' \in U: \; \varphi(x') \cap V \ne \varnothing$.
\end{definition}

Using the same notations as in our definitions, any subset $S$ of $\mathcal{C}(X,Y)$ can be seen as a multi-valued function:
\begin{equation*}
\begin{split}
S : X & \longrightarrow 2^{Y} \\
x & \longmapsto S(x)(=\{f(x) | f \in S\}).
\end{split}
\end{equation*}
Another multi-valued function can be associated to $S$:
\begin{equation*}
\begin{split}
\tilde{S} : X & \longrightarrow 2^{Y} \\
x & \longmapsto \overline{S(x)}.
\end{split}
\end{equation*}

We have the following result:

\begin{propo}\label{C0toCarriers}
Let $S \subset \mathcal{C}(X,Y)$. Then the multi-valued functions $S$ and $\tilde{S}$ are l.h.c.
\end{propo}

\begin{demo}
Let $V$ be an open subset of $Y$, and $V'=\{x \in X | S(x) \cap V \ne \varnothing \}$. We have:
\begin{equation*}
\begin{split}
x \in V' & \Leftrightarrow S(x) \cap V \ne \varnothing \\
& \Leftrightarrow \exists f \in S: \; f(x) \in V \\
& \Leftrightarrow \exists f \in S: \; x \in f^{-1}(V) \\
& \Leftrightarrow x \in \bigcup_{f \in S} f^{-1}(V),
\end{split}
\end{equation*}
thus $V' = \cup_{f \in S} f^{-1}(V)$. Since every $f \in S$ is continuous and $V$ is open, $f^{-1}(V)$ is open, and so is $V'$, which proves that $S$ is l.h.c. 

Given any $V,W \subset Y$ with $V$ is open, we have: $\overline{W} \cap V \ne \varnothing \Leftrightarrow W \cap V \ne \varnothing$. Thus: $\{x \in X | \overline{S(x)} \cap V \ne \varnothing \}=\{x \in X | S(x) \cap V \ne \varnothing \}$, from which one infers that $\tilde{S}$ is l.h.c.
\end{demo}

Let us now go back to our isocones. Let $I \in \mathcal{I}(A)$. We have $I \subset \mathcal{C}(M,\operatorname{Re } M_{n}(\mathbb{C}))$. Therefore, according to Proposition \ref{C0toCarriers}, the local isocone map:

\begin{equation*}
\begin{split}
L : M & \longrightarrow \mathcal{I}_{n} \subset 2^{\operatorname{Re } M_{n}(\mathbb{C})} \\
x & \longmapsto \overline{I(x)}
\end{split}
\end{equation*}  
is a l.h.c. multi-valued function from $M$ into $\operatorname{Re } M_{n}(\mathbb{C})$.

Let $L$ be any multi-valued function $L : M \longrightarrow \mathcal{I}_{n}$. Using the fact that, for all $x \in M$, $L(x)$ is a cone that contains the constant, one can easily prove that:

\begin{equation}
\begin{split}
\mathcal{C}(M,\mathbb{R}) & \subset \mathscr{S}(L) \\
\mathcal{C}(M,\mathbb{R}_{+}) \cdot \mathscr{S}(L) & \subset \mathscr{S}(L).
\end{split}
\label{C+L}
\end{equation}

Starting from an arbitrary local isocone map $L : M \longrightarrow \mathcal{I}_{n}$ that is l.h.c., can one build an isocone? The answer to that question is yes and, as we will see later, this condition is the second half of our necessary and sufficient conditions on lexicographic orders. To prove this, we will need the following theorem (Lemma 5.2 in \cite{Michael}) on selections of l.h.c. multi-valued functions:

\begin{theorem} {\bf (Michael selection theorem)}
Let $X$ be a perfectly normal space\footnote{$X$ is said to be perfectly normal if, for all $E,F$ disjoint non-empty closed subsets of $X$, there exists a function $f \in \mathcal{C}(X, \mathbb{R})$ such that $f^{-1}(\{0\})=E$, and $f^{-1}(\{1\})=F$.}, $Y$ a separable Banach space, and $\varphi : X \longrightarrow 2^{Y}$ a l.h.c. multi-valued function such that $\varphi(x)$ is closed and convex in $Y$ for all $x \in X$. Then there exists a countable subset $\Phi$ of $\mathscr{S}(\varphi)$ such that $\forall x \in X: \; \varphi(x)=\overline{\Phi(x)}$.
\end{theorem}

From this, we deduce (easily) that:

\begin{equation}
\forall x \in X: \; \varphi(x)=\overline{\mathscr{S}(\varphi)(x)}.
\label{SelectionsToCarrier}
\end{equation}

Let us apply this to isocone maps. As stated above, the space $M$ can be assumed metrizable. Therefore, we will restrict ourselves to this case from now on. This assumption is important because it allows us to use the Levin theorem, as well as the Michael selection theorem. Indeed, all metrizable spaces are perfectly normal.

\begin{propo}\label{LHCtoIsocone}
Assume $M$ to be metrizable. Let $L : M \longrightarrow \mathcal{I}_{n}$ be a local isocone map. If $L$ is l.h.c., then $\mathscr{S}(L)$ is an isocone, and:
\begin{equation}
\forall x \in M: \; L(x)=\overline{\mathscr{S}(L)(x)}.
\label{select_dense}
\end{equation}
\end{propo}

\begin{demo}
$M$ is metrizable. It is therefore perfectly normal. We are clearly in a situation where the Michael selection theorem applies. We know, from \eqref{SelectionsToCarrier}, that $\forall x \in M: \; L(x)=\overline{\mathscr{S}(L)(x)}$.

Proving that $\mathscr{S}(L)$ is a pre-isocone is straightforward. To prove that it is an isocone, we need to prove that it generates the algebra. We will start by showing that there exists a finite set $s_{1},...,s_{b}$ of selections of $L$ that generates $\operatorname{Re } M_{n}(\mathbb{C})$ everywhere on $M$. Since $L$ is l.h.c., there exists a countable subset $(\psi_{i})_{i \in \mathbb{N}}$ of its continuous selections, such that $(\psi_{i}(x))_{i \in \mathbb{N}}$ is dense in $L(x)$ for all $x \in M$. Let $x \in M$. There exists a finite subset of $(\psi_{i}(x))_{i \in \mathbb{N}}$ that forms a basis of $\operatorname{Re } M_{n}(\mathbb{C})$. The elements of this finite subset have the indices $n_{1}(x),...,n_{D}(x)$ (with $D=\dim \operatorname{Re } M_{n}(\mathbb{C}) =\frac{n(n-1)}{2}$). The $\psi_{i}$ being continuous mappings, there exists an open neighborhood $V_{x}$ of $x$ such that $(\psi_{n_{i}(x)}(y))_{i=1...D}$ is a basis of $\operatorname{Re } M_{n}(\mathbb{C})$ for all $y \in V_{x}$. The neighborhoods $V_{x}$ (with $x \in M$) form a cover of the compact space $M$. We can therefore extract from it a finite cover of $M$: $V_{x_{1}},...,V_{x_{p}}$, with $p \in \mathbb{N}$, and $x_{1},...,x_{p} \in  M$. It is clear that the finite set of continuous selections $(\psi_{n_{i}(x_{j})})$ (with $i=1...D$, $j=1,...,p$) generates $\operatorname{Re } M_{n}(\mathbb{C})$ everywhere on $M$. Let us denote its elements: $s_{1},...,s_{b} \in \mathscr{S}(L)$ ($b=pD$).

Let $F \in \mathcal{C}(M,\operatorname{Re } M_{n}(\mathbb{C}))$. There exists (not necessarily continuous) mappings $f_{i}:M \rightarrow \mathbb{R}$, $i=1,...,b$, such that $F=\sum_{i} f_{i} s_{i}$. All we need to do now is prove that these mappings can be chosen continuous. Let $\vec{f}$ be the $b$-dimensional vector that contains the $f_{i}$, $\vec{F}$ the $D$-dimensional vector that contains the components of $F$ in the canonical basis of $\operatorname{Re } M_{n}(\mathbb{C})$, and $A$ the $D \times b$ real matrix that has the $s_{i}$ as columns when expressed in the canonical basis. We have $\vec{F}=A \vec{f}$. Since the $s_{i}$ generate the real vector space $\operatorname{Re } M_{n}(\mathbb{C})$ at all points, $A$ is invertible from the left everywhere; there exists a (not necessarily continuous) real mapping $B : M \rightarrow M_{b,D}(\mathbb{R})$ such that $BA=I_{b}$ on $M$. The $f_{i}$ are now simply given by $\vec{f}=B \vec{F} $. We see that in order to have $\vec{f}$ continuous, we need to choose $B$ continuous. To this end, we will use the Michael selection theorem again.

Let us define the multi-valued function:
\begin{equation*}
\begin{split}
\mathcal{B}: M & \longrightarrow 2^{M_{b,D}(\mathbb{R})} \\
x & \longmapsto \mathcal{B}(x)=\{B \in M_{b,D}(\mathbb{R}) | B \times A(x)=I_{b}\}.
\end{split}
\end{equation*}
For all $x \in M$, $\mathcal{B}(x)$ is a finite dimensional affine space. It is therefore closed and convex. Let us now prove that $\mathcal{B}$ is lower hemi-continuous. Let $x \in M$, and $B \in \mathcal{B}(x)$. We have by definition of $\mathcal{B}$ that $BA(x)=I_{b}$. Let $V$ be an open neighborhood of $B$ in $M_{b,D}(\mathbb{R})$, and $J: y \mapsto BA(y)$. The map $J$ is continuous since $A$ is, and $J(x)=I_{b}$. Therefore, there exists a neighborhood $U_{1}$ of $x$ such that for all $y \in U_{1}$ we have $\left\| J(y)-I_{b} \right\| < 1$. Thus, $J(y)$ is invertible. The mapping $y \longmapsto J(y)^{-1}$ is obviously well-defined and continuous on $U_{1}$, and so is $y \longmapsto J(y)^{-1}B$. Since we have $J(x)^{-1}B=B$, there exists a neighborhood $U \subset U_{1}$ of $x$ such that $J(y)^{-1}B \in V$ for all $y \in U$. Notice that $J(y)^{-1}B \times A(y)=I_{b}$ by definition, and thus that $J(y)^{-1}B \in \mathcal{B}(y)$. We just proved that for all $x \in M$, $B \in \mathcal{B}(x)$, and $V$ neighborhood of $B$, there exists a neighborhood $U$ of $x$ such that for all $y \in U$ : $\mathcal{B}(y) \cap V \ne \varnothing$. This is just the statement that $\mathcal{B}$ is l.h.c.. We conclude that it has continuous selections. Let $\tilde{B}$ be such a continuous selection. Taking $\vec{f}=\tilde{B} \vec{F} $, we now have continuous components for the element $F$: $f_{i} \in \mathcal{C}(M, \mathbb{R})$. Each $f_{i}$ can be written as the difference of two positive continuous functions: $f_{i}=f_{i,+}-f_{i,-}$ (\emph{e.g.} $f_{i,+}=\max(0,f), f_{i,-}=\min(0,-f)$). Therefore, we have $F=F_{+}-F_{-}$, where $F_{\pm}=\sum_{i} f_{i,\pm} s_{i} \in \mathscr{S}(L)$ (thanks to \eqref{C+L}). Thus $\mathcal{C}(M,\operatorname{Re } M_{n}(\mathbb{C})) \subset \mathscr{S}(L)-\mathscr{S}(L)$, which implies that $\mathcal{C}(M,\operatorname{Re } M_{n}(\mathbb{C})) = \mathscr{S}(L)-\mathscr{S}(L)$. This concludes our proof.
\end{demo}

One can prove, using \eqref{C+L}, that the order induced by the isocone $\mathscr{S}(L)$ is the lexicographic combination of the trivial order on $M$, and the local order induced by the map $L$.

\subsection{Existence of isocones inducing a lexicographic order}

We can now state the necessary and sufficient conditions we are looking for.

\begin{theorem}\label{Class MnC 2}
Assume $M$ to be metrizable. Let $\preceq_{M}$ be an order on $M$, and $L : M \longrightarrow \mathcal{I}_{n}$ be a local isocone map. There exists an isocone $I \in \mathcal{I}(A)$ such that, for all $(x,\xi),(y,\eta) \in P(A)$:
\begin{equation}
(x,\xi) \leq_{I} (y,\eta) \Leftrightarrow 
\begin{cases}
x \prec_{M} y \\
\text{or} \\
x=y \text{ and } \xi \preceq_{L(x)} \eta,
\end{cases}
\label{lexico3}
\end{equation}
if and only if the following requirements are met:
\begin{enumerate}
\item $\prec_{M}$ is closed,
\item $L$ is l.h.c..
\end{enumerate}
\end{theorem}

\begin{demo}

We already know from the preceding sections that these conditions are necessary. Let us prove that they are sufficient. For this purpose, consider the following subset of $\operatorname{Re } A$:

\begin{equation}
\begin{split}
I =& \operatorname{Lex }_{x \in (M,\preceq_{M})} L(x) \\
 :=& \{F \in \mathscr{S}(L) | \forall x,y \in M: \; x \prec_{M} y \Rightarrow \max \sigma(F(x)) \leq \min \sigma(F(y)) \}.
\end{split}
\label{Isolexico}
\end{equation}

We will first prove that $I$ is an isocone. Then we will prove that the order it induces is given by \eqref{lexico3}. Notice first that, thanks to Proposition \ref{LHCtoIsocone}, $\mathscr{S}(L)$ is an isocone. 

It is clear that $I$ is closed and contains $\mathbb{R}1_{A}$. Now, let $F,G \in I$, $\phi \in I(\mathbb{R})$ and $x,y \in M$ such that $x \prec_{M} y$. We have:

\begin{equation*}
\begin{split}
\max \sigma((F+G)(x)) & \leq \max \sigma(F(x))+ \max \sigma(G(x))\\
& \leq \min \sigma(F(y))+ \min \sigma(G(y)) \\
& \leq \min \sigma((F+G)(y)) \\
\end{split},
\end{equation*}
and $F+G \in \mathscr{S}(L)$, Therefore $F+G \in I$. We also have:

\begin{equation*}
\begin{split}
\max \sigma(\phi \circ F(x)) & = \max \sigma(\phi(F(x))) \\
& = \max \phi(\sigma(F(x))) \\
& = \phi(\max \sigma(F(x))) \text{, since $\phi$ is increasing} \\
& \leq \phi(\min \sigma(F(y))) \text{, since $F \in I$ and $\phi$ is increasing} \\
& \leq \min \sigma(\phi \circ F(y)) \text{, by the same steps as above,}
\end{split}
\end{equation*}
and $\phi \circ F \in \mathscr{S}(L)$. Therefore $\phi \circ F \in I$. Thus $I$ is a pre-isocone.

Now, let $F \in \mathscr{S}(L)$. Let $m = \inf_{x \in M} \min \sigma(F(x))$. Then $F'=F-m \in \mathscr{S}(L)$ is positive. Define $\Lambda=\sup_{x \in M} \max \sigma(F'(x))$. According to Corollary \ref{Levin}, $(M, \preceq_{M})$ is a compact toposet and there exists a strictly increasing continuous real function $g$ on $M$ such that:

\begin{equation}
\forall x,y \in M: \; x \prec_{M} y \Rightarrow g(y)-g(x) \geq 1.
\end{equation}
Notice that, by the definition of $I$ and \eqref{C+L}:

\begin{equation*}
I(M,\preceq_{M}) \subset I.
\end{equation*}

Thus $g \in I$. Consider $G=F'+\Lambda g \in \mathscr{S}(L)$. For all $x,y \in M$ such that $x \prec_{M} y$, we have:

\begin{equation*}
\begin{split}
\min \sigma(G(y)) & = \min \sigma(F'(y)) + \Lambda g(y) \\
& \geq \Lambda g(x) + \Lambda \text{, since $F'$ is positive, and by definition of $g$} \\
& \geq \Lambda g(x) + \max \sigma(F'(x)) \\
& \geq \max \sigma(G(x)).
\end{split}
\end{equation*}
Thus $G \in I$. Consequently: $F=G - \lambda g + m \in I-I$. This proves that $\mathscr{S}(L) \subset I-I$, and thus that $\mathscr{S}(L)-\mathscr{S}(L) \subset I-I$. We know that $\mathscr{S}(L)-\mathscr{S}(L)=\operatorname{Re } A$. Hence $\overline{I-I}=\operatorname{Re } A$. Thus $I$ is an isocone.

This implies the existence of an order $\preceq_{M}^{I}$ on $M$ such that, for all $(x,\xi),(y,\eta) \in P(A)$:

\begin{equation}
(x,\xi) \leq_{I} (y,\eta) \Leftrightarrow 
\begin{cases}
x \prec_{M}^{I} y \\
\text{or} \\
x=y \text{ and } \xi \preceq_{\overline{I(x)}} \eta.
\end{cases}
\end{equation}
The order $\preceq_{M}^{I}$ is given by \eqref{M_strictorder}:

\begin{equation}
\forall x,y \in M: \; x \prec_{M}^{I} y \Leftrightarrow \forall F \in I: \; \max \sigma(F(x)) \leq \min \sigma(F(y)).
\end{equation}
Let $x,y \in M$. It is obvious that if $x \prec_{M} y$, then $x \prec_{M}^{I} y$. Conversely, suppose that $x \prec_{M}^{I} y$. Then we have $\forall F \in I: \; \max \sigma(F(x)) \leq \min \sigma(F(y))$. We know that $I(M,\preceq_{M}) \subset I$. Thus: $\forall f \in I(M,\preceq_{M}): \; f(x) \leq f(y)$, hence $x \preceq_{M} y$. The points $x$ and $y$ are distinct because $x \prec_{M}^{I} y$. Thus $x \prec_{M} y$. One concludes that $\prec_{M}^{I}$ and $\prec_{M}$ coincide, and so do $\preceq_{M}^{I}$ and $\preceq_{M}$.

All that remains to prove is that $\overline{I(x)}=L(x)$ for all $x \in M$. According to Theorem \ref{Class MnC 1}, we have, for $x \in M$: $L(x)=\overline{\mathscr{S}(L)(x)}$. Let $x \in M$. From $I \subset \mathscr{S}(L)$, we find $I(x) \subset \mathscr{S}(L)(x)$.  Conversely, Let $A \in \mathscr{S}(L)(x)$. Then there exists $F \in \mathscr{S}(L)$ such that $A = F(x)$. Let $\mu < \inf_{y \in M} \min \sigma(F(y))$. Then $F' = F - \mu \in \mathscr{S}(L)$ is strictly positive. Let $\Omega \geq 0$ be defined by:

\begin{equation*}
e^{\Omega}=\frac{\sup_{y \in M} \max \sigma(F'(y))}{\inf_{y \in M} \min \sigma(F'(y))} \geq 1,
\end{equation*}
and $H = e^{\Omega g} F' \in \mathscr{S}(L)$ (to see that it is in $\mathscr{S}(L)$, use \eqref{C+L}). This is clearly a strictly positive element of the algebra. For $y,z \in M$ such that $y \prec_{M} z$, we have:

\begin{equation*}
\begin{split}
\frac{\min \sigma(H(z))}{\max \sigma(H(y))} & = e^{\Omega (g(z)-g(y))} \frac{\min \sigma(F'(z))}{\max \sigma(F'(y))} \\
& \geq e^{\Omega} \frac{\inf_{y \in M} \min \sigma(F'(y))}{\sup_{y \in M} \max \sigma(F'(y))} \\
& \geq e^{\Omega} e^{-\Omega}=1.
\end{split}
\end{equation*}
Thus $H \in I$. We have $A=F(x)=\mu+e^{-\Omega g(x)}H(x) \in I(x)$ (it is easy to prove that $I(x)$ is a convex cone that contains the constants). Hence $\mathscr{S}(L)(x) \subset I(x)$. We thus conclude that $I(x) = \mathscr{S}(L)(x)$. Taking the closure, one finds that $\overline{I(x)} = L(x)$. This completes our proof.
\end{demo}

We now have completed the classification of isocone-induced orders when $M$ is metrizable for the particular case of the algebra $A=\mathcal{C}(M) \otimes M_{n}(\mathbb{C})$.

\subsection{Application to Lorentzian manifolds}

Let us now give an example of an order satisfying the conditions of Theorem \ref{Class MnC 2} on a Lorentzian manifold. Consider first the $D$-dimensional flat spacetime $\mathbb{R}^{1,D-1}$ with metric signature $(+ - \ldots -)$.  The Lorentzian causal order is given by:
\begin{equation*}
\forall x,y \in \mathbb{R}^{1,D-1}: \; x \preceq y \Leftrightarrow \begin{cases}
x^{0} \leq y^{0} \\
\text{and} \\
(y-x)^{2} \geq 0.
\end{cases}
\end{equation*}
This order does not satisfy the property that $\prec$ is closed. Therefore, we must look for an order that satisfies this property, is orthocronous-Poincar\'{e} invariant, and approximates Lorentzian order. For $\Lambda \geq 0$, we define the following subsets of $\mathbb{R}^{1,D-1}$:

\begin{equation*}
\begin{split}
\mathscr{C}(\Lambda) & = \{x | x^{0} \geq 0 \text{ and } x^{2} \geq \Lambda^{2} \} \\
\mathscr{C}^{\calligra{o}}(\Lambda) & = \mathscr{C}(\Lambda) \cup \{0\}.
\end{split}
\end{equation*}
The order we are looking for is:

\begin{equation}
x \preceq_{\Lambda} y \Leftrightarrow y-x \in \mathscr{C}^{\calligra{o}}(\Lambda),
\label{lambdaorder}
\end{equation}
with $\Lambda > 0$. Indeed, we have:

\begin{equation*}
x \prec_{\Lambda} y \Leftrightarrow y-x \in \mathscr{C}(\Lambda),
\end{equation*}
which is a closed relation, since the set $\mathscr{C}(\Lambda)$ is closed. Notice that for $\Lambda = 0$, the order $\preceq_{\Lambda}$ simply reduces to the Lorentzian order. This order can be generalized to causal spacetimes. For this, we need a Lorentzian distance on $M$, that is a distance $d$ such that $d(x,y)$ is the length of the geodesic connecting $x$ and $y$ if they are causally related (in the Lorentzian sense), and vanishes if $x$ and $y$ are not causally related. In flat spacetime, this is simply $d(x,y)=\sqrt{\max(0,(x-y)^{2})}$. The order $\preceq_{\Lambda}$ is now defined by:
\begin{equation*}
x \prec_{\Lambda} y \Leftrightarrow \begin{cases}
x^{0} \leq y^{0} \\
\text{and} \\
d(x,y) \geq \Lambda,
\end{cases}
\end{equation*}
where $\Lambda > 0$. This order is obviously invariant by all orthocronous diffeomorphisms, and approximates Lorentzian order for small values of $\Lambda$.

\begin{figure}
\includegraphics[width=\columnwidth]{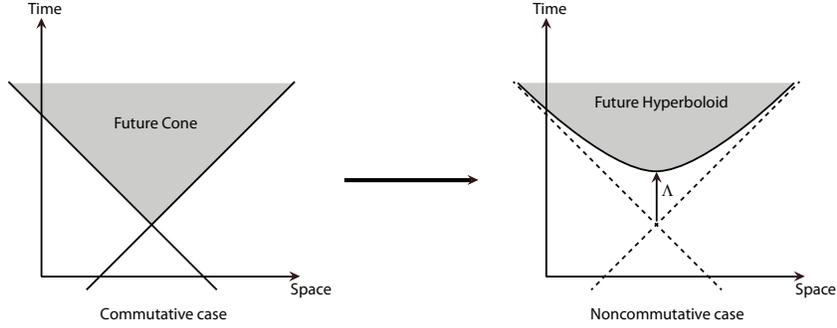}
\caption{Lorentzian causality needs to be altered on noncommutative manifolds.}
\label{FutureHyperboloid}
\end{figure}

\subsection{General finite-dimensional algebra}

We know consider the algebra $A=\mathcal{C}(M) \otimes A_{f}$. We give the classification theorems for such an algebra, their proofs are delayed until the appendix.

\begin{theorem}\label{Class Af 1} Let $I \in \mathcal{I}(A)$. There exists an order $\preceq_{M_{K}}$ on $M_{K}$ such that for all $(k,x,\xi),(l,y,\eta) \in P(A)$:
\begin{equation}
(k,x,\xi) \leq_{I} (l,y,\eta) \Leftrightarrow 
\begin{cases}
(k,x) \prec_{M_{K}} (l,y) \\
\text{or} \\
(k,x)=(l,y) \text{ and } \xi \leq_{I_{k,x}} \eta,
\end{cases}
\label{lexico4}
\end{equation}
where $I_{k,x}=\overline{\pi_{k}(I(x))} \in \mathcal{I}_{n_{k}}$.
\end{theorem}

Let us define:

\begin{equation}
\mathscr{G}_{k,l}=\{(x,y) \in M \times M | (k,x) \prec_{M_{K}} (l,y)\}. 
\label{klgraph}
\end{equation}

\begin{theorem}\label{Class Af 2}
Assume $M$ to be metrizable. Let $\preceq_{M_{K}}$ be an order on $M_{K}$, and $L_{k} : M \longrightarrow \mathcal{I}_{n_{k}},k=1..K$ be a set of local isocone maps. There exists an isocone $I \in \mathcal{I}(A)$ such that, for all $(k,x,\xi),(l,y,\eta) \in P(A)$:
\begin{equation}
(k,x,\xi) \leq_{I} (l,y,\eta) \Leftrightarrow 
\begin{cases}
(k,x) \prec_{M_{K}} (l,y) \\
\text{or} \\
(k,x)=(l,y) \text{ and } \xi \preceq_{L_{k}(x)} \eta,
\end{cases}
\label{lexico6}
\end{equation}
if and only if the following requirements are met:
\begin{enumerate}
\item $\preceq_{M}$ is closed, and for all $k,l \in \llbracket 1,K \rrbracket$ such that $(n_{k},n_{l}) \ne (1,1)$, $\mathscr{G}_{k,l}$ is closed in $M \times M$.
\item for all $k \in \llbracket 1,K \rrbracket$, $L_{k}$ is l.h.c.
\end{enumerate}
\end{theorem}

We see that the order locally vanishes on the copies of M which are associated to a noncommutative algebra. Moreover, the construction of examples of orders meeting the requirements of the theorem and approximating the usual causal order on spacetime gives rise to interesting combinatorial conditions (see the appendix).

\section{Discussion}

A few words need to be said about the physical consequences of our theorems. For the sake of simplicity, we will focus on the simple case described in Theorem \ref{Class MnC 2}. We assume that $M$ is metric, with a distance $d$ defined on it. Let $I \in \mathcal{I}(A)$, and $\preceq_{M}$ the order it induces on $M$. Since the set $\{(x,y) \in M \times M | x \prec_{M} y \}$ is closed, it is compact. The distance $d$ is continuous, therefore it reaches its infimum on that set, hence:

\begin{equation}
\varepsilon_{0}=\inf_{x \prec_{M} y} d(x,y) > 0. 
\label{cutoff}
\end{equation}
By definition of $\varepsilon_{0}$, we have:

\begin{equation}
\forall x \ne y: \; d(x,y) < \varepsilon_{0} \Rightarrow x \|_{M} y.
\label{orderlimit1}
\end{equation}
In other words, \emph{there exists a scale below which the order ceases to exist}. This scale is a global characteristic of $M$: for a given isocone, the order on $M$ cannot be arbitrarily fine. The manifold $M$ can better be thought of as a compact neighbourhood in a given spacetime. The requirement \eqref{orderlimit1} is obviously met when $M$ is a compact, hence finite, piece of a discrete manifold. As such, our results can be understood as a motivation for discrete approaches to quantum gravity. It is however interesting to note that discreteness is not imposed by our theorem, in fact we have given above an example of order relation defined on a smooth manifold and satisfying \eqref{orderlimit1}.  

The absence of order at small distances is only true in the noncommutative case ($n \geq 2$). This shows that the noncommutativity of the algebra greatly constrains the allowed orders. More surprisingly, the reverse is true as well. In the case $n \geq 3$, all local isocones are trivial and the local order is as coarse as possible. But for $n=2$, the local order is not necessarily trivial. In that case, let us build the multi-valued function $K : M \longrightarrow 2^{S^2}$ that associates a subset of the sphere to the local isocone $\overline{I(x)}$ such that $\overline{I(x)}=\mathbb{R}_{+}K(x)+\mathbb{R}1_{2}$. It is easy to see that $K$ is l.h.c. Let $\mu : 2^{S^2} \longrightarrow \mathbb{R}_{+}$ be a strictly positive measure of the sphere - that is, $\mu$ is strictly positive for all open subsets of $S^2$. Then one can prove that $a : x \longmapsto \mu(K(x)) \in \mathbb{R}^{M}$ is lower semi-continuous. Given that $M$ is compact, $a$ reaches its infimum on it: there exists $x_{0} \in M$ such that: $a_{0} \equiv \inf_{M} a = a(x_{0}) =  \mu(K(x_{0}))$. Because $K(x_{0})$ is of non-empty interior, $a_{0}$ is strictly positive. We therefore have:

\begin{equation}
\forall x \in M: \; \mu(K(x)) \geq a_{0} > 0.
\label{orderlimit2}
\end{equation}
Thus the local isocones cannot be arbitrarily narrow; this is the non-closing cones phenomenon (see \cite{bes3}). In other words, \emph{the `internal' order cannot be arbitrarily fine}. This proves that the commutative component of the algebra influences the noncommutative one as well. It is, in both cases, the manifestation of the compatibility between topology and order that we have imposed.

The disappearance of causality at small scales is remarkably coherent with the physical properties of the spectral action, despite the fact that it does not appear at any point in isocone theory. Indeed, it is shown in \cite{lizzi} and \cite{asz} that bosons do not propagate at high energies, and thus at small scales. The contribution of the high energy components of their propagators is shown to be proportional to a delta distribution. One wonders whether this indicates a hidden connection between causality and the spectral action. This might also be an indication that the almost-commutative manifold hypothesis breaks down at high energies, where quantum gravity effects are not negligible anymore.

\section*{Acknowledgments}

We would like to thank Ettore Minguzzi for his help in carefully distinguishing the different causality conditions on Lorentzian manifolds. We would also like to thank Christian Brouder for the interesting physical and mathematical discussions and the useful advice.

\appendix

\section{Proof of the classification theorems for a general almost-commutative algebra} 

\subsection{Proof of Theorem \ref{Class Af 1}}

\begin{demo}
The proof is similar to that of Theorem \ref{Class MnC 1}. We will thus not give it in details.

Let $P$ be a finite subset of $M$. We denote $P_{K}=\llbracket 1,K \rrbracket \times P \subset M_{K}$. Consider $ev_{P}(A)=\bigoplus_{k=1}^{K} (A_{f})_{x} \cong \bigoplus_{(k,x) \in P_{K}} M_{n_{k}}(\mathbb{C})_{x}$ the restriction of $A$ to $P$. Its pure state space is $P(ev_{P}(A)) \cong \coprod_{k=1}^{K} (P \times  \CC P^{n_{k}-1}) \subset P(A)$. We denote $\pi_{k,x}$ the projection on the $M_{n_{k}}(\mathbb{C})_{x}$ component of the restricted algebra. Let $I \in \mathcal{I}(A)$. Then $I_{P} \equiv \overline{ev_{P}(I)} \in \mathcal{I}(ev_{P}(A))$. Similarly to what is done in Theorem \ref{Class MnC 1}, on can prove that $\leq_{I}$ and $\leq_{I_{P}}$ coincide on $P$.

$I_{P}$ is finite-dimensional. Therefore, there exists an order $\preceq_{P_{K}}$ on $P_{K}$ such that for all $(k,x,\xi),(l,y,\eta) \in P(ev_{P}(A))$:

\begin{equation}
(k,x,\xi) \leq_{I} (l,y,\eta) \Leftrightarrow 
\begin{cases}
(k,x) \prec_{P_{K}} (l,y) \\
\text{or} \\
(k,x)=(l,y) \text{ and } \xi \leq_{I_{k,x}} \eta,
\end{cases}
\label{lexico5}
\end{equation}
where $I_{k,x}=\pi_{k,x}(I_{P}) \in \mathcal{I}_{n_{k}}$. One can prove that $I_{k,x}=\overline{\pi_{k}(I(x))}$.

Let $(\xi_{k})_{k=1..K}$ be a collection of fixed elements such that $\xi_{k} \in \CC P^{n_{k}-1}$. Consider the binary relation $\preceq_{M_{K}}$ on $M_{K}$ defined by:

\begin{equation}
\forall (k,x),(l,y) \in M_{K}: \; (k,x) \preceq_{M_{K}} (l,y) \Leftrightarrow (k,x,\xi_{k}) \leq_{I} (l,y,\xi_{l}).
\label{MKQ_order}
\end{equation}
This is a closed order on $M_{K}$. Using \eqref{lexico5}, one easily proves that $\preceq_{P_{K}}$ and $\preceq_{M_{K}}$ coincide on $P_{K}$. We rewrite \eqref{lexico5} as:

\begin{equation*}
\forall (k,x,\xi),(l,y,\eta) \in P(ev_{P}(A)), \; (k,x,\xi) \leq_{I} (l,y,\eta) \Leftrightarrow
\begin{cases}
(k,x) \prec_{M_{K}} (l,y) \\
\text{or} \\
(k,x)=(l,y) \text{ and } \xi \leq_{I_{k,x}} \eta.
\end{cases}
\end{equation*}
This is once again true for all $P$ finite subset of $M$. Hence our theorem.
\end{demo}

We see that $\preceq_{M_{K}}$ is closed. Therefore $(M_{K}, \preceq_{M_{K}})$ is a compact toposet. We now prove that, for $k,l \in \llbracket 1,K \rrbracket$ such that $(n_{k},n_{l}) \ne (1,1)$:

\begin{equation}
\forall x,y \in M: \; (k,x) \prec_{M_{K}} (l,y) \Leftrightarrow \forall F \in I: \; \max \sigma(F_{k}(x)) \leq \min \sigma(F_{l}(y)).
\label{MK_strictorder}
\end{equation}

Let $k,l \in \llbracket 1,K \rrbracket$ be such that $(n_{k},n_{l}) \ne (1,1)$, and $x,y \in M$. If $(k,x) \prec_{M_{K}} (l,y)$, then $(k,x,\xi) \leq_{I} (l,y,\eta)$ for all $\xi \in \CC P^{n_{k}-1}, \eta \in \CC P^{n_{l}-1}$. Conversely, suppose that $(k,x,\xi) \leq_{I} (l,y,\eta)$ for all $\xi \in \CC P^{n_{k}-1}, \eta \in \CC P^{n_{l}-1}$. Then $(k,x) \preceq_{M_{K}} (l,y)$. If $(k,x)=(l,y)$, then for all $\xi \in \CC P^{n_{k}-1}, \eta \in \CC P^{n_{l}-1}: \; \xi \leq_{I_{k,x}} \eta$, which is absurd (since here $n_{k}=n_{l} \ne 1$). Therefore $(k,x)$ and $(l,y)$ are distinct, and we have: $(k,x) \prec_{M_{K}} (l,y)$. Thus:

\begin{equation*}
\begin{split}
(k,x) \prec_{M_{K}} (l,y) & \Leftrightarrow \forall \xi \in \CC P^{n_{k}-1}, \eta \in \CC P^{n_{l}-1}: \; (k,x,\xi) \leq_{I} (l,y,\eta) \\
& \Leftrightarrow \forall F \in I: \; \max \sigma(F_{k}(x)) \leq \min \sigma(F_{l}(y)).
\end{split} 
\end{equation*}

From \eqref{MK_strictorder}, we conclude that for all $k,l \in \llbracket 1,K \rrbracket$ such that $(n_{k},n_{l}) \ne (1,1)$, the graph:

\begin{equation*}
\mathscr{G}_{k,l}=\{(x,y) \in M \times M | (k,x) \prec_{M_{K}} (l,y)\} 
\end{equation*}
is closed. We now have the necessary and sufficient conditions on $\preceq_{M_{K}}$.

Consider the local isocone maps:

\begin{equation*}
\begin{split}
L_{k} : M & \longrightarrow \mathcal{I}_{n_{k}} \\
x & \longmapsto I_{k,x}=\overline{\pi_{k}(I(x))}.
\end{split}
\end{equation*} 
Let $I_{k}$ be the projection of $I$ on the $k$-th component:
\begin{equation*}
I_{k}=\pi_{k}(I) \equiv \{F_{k} | F \in I\} \subset \mathcal{C}(M, \operatorname{Re } M_{n_{k}}(\mathbb{C})).
\end{equation*}
We have, for $x \in M: \pi_{k}(I(x))=I_{k}(x)$. Therefore $L_{k}(x)=\overline{I_{k}(x)}$. Thus the $L_{k}$ are all l.h.c.

\subsection{Proof of Theorem \ref{Class Af 2}}

\begin{demo}

Let $\mathscr{L}=\bigoplus_{k} \mathscr{S}(L_{k})$. According to Proposition \ref{LHCtoIsocone}, $\mathscr{S}(L_{k})$ is an isocone for every $k=1...K$. It is not hard to see that $\mathscr{L}$ is an isocone of $A$. Moreover, \eqref{C+L} can be generalized here as:

\begin{equation*}
\begin{split}
\mathcal{C}(M_{K},\mathbb{R}) & \subset \mathscr{L} \\
\mathcal{C}(M_{K},\mathbb{R}_{+}) \cdot \mathscr{L} & \subset \mathscr{L},
\end{split}
\end{equation*}
where we used the embedding of $\mathcal{C}(M_{K})$ in $A$ described in section (3.1).

Consider the following subset of $\operatorname{Re } A$:

\begin{equation}
I = \{F \in \mathscr{L} | (k,x) \prec_{M_{K}} (l,y) \Rightarrow \max \sigma(F_{k}(x)) \leq \min \sigma(F_{l}(y)) \}.
\label{IsolexicoF}
\end{equation}
Proving that $I$ is a pre-isocone is straightforward and is done in the same fashion as in the proof of Theorem \ref{Class MnC 2}. Let us now prove that it is an isocone.

Notice first that, since $\preceq_{M_{K}}$ is closed, $(M_{K}, \preceq_{M_{K}})$ is a compact toposet. We thus have $\mathcal{C}(M_{K},\mathbb{R}) = \overline{I(M_{K}, \preceq_{M_{K}})-I(M_{K}, \preceq_{M_{K}})}$. It is easy to see that $I(M_{K}, \preceq_{M_{K}}) \subset I$. Hence $\mathcal{C}(M_{K},\mathbb{R}) \subset \overline{I-I}$. Thanks to the Levin theorem, we know that there exists a strictly increasing continuous function $v$ on $M_{K}$. For $k,l=1...K$ such that $(n_{k},n_{l}) \ne (1,1)$, we define:

\begin{equation*}
m_{k,l}=\begin{cases}
\displaystyle\inf_{(x,y) \in \mathscr{G}_{k,l}} (v_{l}(y)-v_{k}(x)) &\text{if } \mathscr{G}_{k,l} \ne \varnothing \\
1 & \text{otherwise.}
\end{cases}
\end{equation*}
Using the compactness of $\mathscr{G}_{k,l}$, we get $m_{k,l} > 0$. Let $m>0$ be the smallest of the $m_{k,l}$. By their definition, the function $g=\displaystyle\frac{v}{m}$ is such that, for all $k,l$ such that $(n_{k},n_{l}) \ne (1,1)$, and all $x,y \in M$:

\begin{equation*}
(k,x) \prec_{M_{K}} (l,y) \Rightarrow g_{l}(y)-g_{k}(x) \geq 1.
\end{equation*}

Now, let $F \in \mathscr{L}$. We can suppose without any loss in generality that $F$ is positive. We decompose it as $F=F^{c}+F^{nc}$, where:

\begin{equation*}
\begin{split}
F^{c}_{k} & = F_{k} \delta_{n_{k},1} \\
F^{nc}_{k} & = F_{k} (1-\delta_{n_{k},1});
\end{split}
\end{equation*}
$F^{c}$ contains the components of $F$ that take their values in the commutative parts of the finite algebra, whereas $F^{nc}$ contains the ones that take their values in the noncommutative parts. Both parts are positive. We obviously have $F^{c} \in \mathcal{C}(M_{K}, \mathbb{R}) \subset \overline{I-I}$. Once again, we define $\Lambda=\sup_{x \in M} \max \sigma(F^{nc}(x))$, and from there: $G=F^{nc}+\Lambda g \in \mathscr{L}$. Let $(k,x) \prec_{M_{K}} (l,y)$. If $(n_{k}, n_{l}) \ne (1,1)$, then:

\begin{equation*}
\begin{split}
\min \sigma(G_{l}(y)) & = \min \sigma(F^{nc}_{l}(y)) + \Lambda g_{l}(y) \\
& \geq \Lambda g_{k}(x) + \Lambda \\
& \geq \Lambda g_{k}(x) + \max \sigma(F^{nc}_{k}(x)) \\
& \geq \max \sigma(G_{k}(x)).
\end{split}
\end{equation*}
If $n_{k}=n_{l}=1$, then $F^{nc}_{k}=F^{nc}_{l}=0$. Thus: $\min \sigma(G_{l}(y)) = \Lambda g_{l}(y) \geq \Lambda g_{k}(x) = \max \sigma(G_{k}(x))$, since $g$ is non-decreasing.
Hence $G \in I$. We have: $F=F^{c}+G- \Lambda g \in I-I +\overline{I-I} \subset \overline{I-I}$. Thus $\mathscr{L} \subset \overline{I-I}$. This implies that $\overline{I-I} =\operatorname{Re } A$. Hence $I$ is an isocone.

We conclude that there exists an order $\prec_{M_{K}}^{I}$ on $M_{K}$ such that:

\begin{equation}
(k,x,\xi) \leq_{I} (l,y,\eta) \Leftrightarrow 
\begin{cases}
(k,x) \prec_{M_{K}}^{I} (l,y) \\
\text{or} \\
(k,x)=(l,y) \text{ and } \xi \preceq_{\overline{I_{k}(x)}} \eta.
\end{cases}
\end{equation}

Let $(k,x) \prec_{M_{K}}^{I} (l,y)$. Then for all $\xi, \eta$: $(k,x,\xi) \leq_{I} (l,y,\eta)$. Thus, for all $F \in I$, we have $\max \sigma(F_{k}(x)) \leq \min \sigma(F_{l}(y))$. Using $I(M_{K}, \preceq_{M_{K}}) \subset I$, we infer that $(k,x) \prec_{M_{K}} (l,y)$. Conversely, $(k,x) \prec_{M_{K}} (l,y)$ implies that for all $F \in I$, we have $\max \sigma(F_{k}(x)) \leq \min \sigma(F_{l}(y))$ (by the definition of $I$). Thus, for all $\xi, \eta$: $(k,x,\xi) \leq_{I} (l,y,\eta)$. Since the order is lexicographic, this implies that $(k,x) \preceq_{M_{K}}^{I} (l,y)$, and thus that $(k,x) \prec_{M_{K}}^{I} (l,y)$. We conclude that the orders $\prec_{M_{K}}^{I}$ and $\prec_{M_{K}}$ coincide.

Finally, let $k \in \llbracket 1,K \rrbracket$, and $x \in M$. We have $I_{k} \subset \mathscr{S}(L_{k})$, hence $I_{k}(x) \subset \mathscr{S}(L_{k})(x)$. Conversely, let $B \in \mathscr{S}(L_{k})(x)$. If $n_{k} = 1$, then $L_{k}(x)=\mathbb{R}$. Hence $B \in \mathbb{R}$. Let $\tilde{B}= B \cdot 1_{A} \in I$. We have $\tilde{B}_{k}(x)=B$. Thus $B \in I_{k}(x)$. If $n_{k} \geq 2$, define $\tilde{B} \in A_{f}$ such that $\tilde{B}_{l}=B \delta_{lk}$. It is clear that $\tilde{B} \in \mathscr{L}(x)$. Using the same method as in the proof of Theorem \ref{Class MnC 2}, one can prove that $\tilde{B} \in I(x)$, and thus that $B \in I_{k}(x)$. We conclude that $\mathscr{S}(L_{k})(x) \subset I_{k}(x)$, from which one can infer that $I_{k}(x) = \mathscr{S}(L_{k})(x)$. Taking the closure, we find: $L_{k}(x) = \overline{I_{k}(x)}$.
\end{demo}

\subsection{Properties and an example}

The properties described in section (4) apply here. Indeed, the local isocones cannot be arbitrarily narrow. Moreover, if $d$ is the distance on $M$, then for all $k$ such that $n_{k} \geq 2$, there exists a scale $\varepsilon_{k} > 0$ such that:

\begin{equation}
\forall x \ne y: \; d(x,y) < \varepsilon_{k} \Rightarrow (k,x) \|_{M_{K}} (k,y)
\end{equation}
(this can be proved using that $\mathscr{G}_{kk}$ is closed). That is, order ceases to exists at small scales on the noncommutative copies of $M$ only.

To conclude this appendix, let us give an example of such an order; we consider a $D$-dimensional flat spacetime $\mathbb{R}^{1,D-1}$, and we want to generalize Lorentzian causality as in the example of section (3.5). We are looking for an order that is orthocronous-Poincar\'{e} invariant, under the assumption that all the copies of $M$ in $M_{K}$ transform simultaneously under a Poincar\'{e} transformation. For this purpose, we consider a partition $P,P^{\calligra{o}}$ of $\llbracket 1,K \rrbracket^{2}$. Then one can define the binary relation $\preceq_{\Lambda}$ such that:

\begin{equation}
\begin{split}
\forall x,y \in M, \forall (k,l) \in P: \; (k,x) \preceq_{\Lambda} (l,y) &\Leftrightarrow y-x \in \mathscr{C}(\Lambda_{k,l}), \\
\forall x,y \in M, \forall (k,l) \in P^{\calligra{o}}: \; (k,x) \preceq_{\Lambda} (l,y) &\Leftrightarrow y-x \in \mathscr{C}^{\calligra{o}}(\Lambda_{k,l}),
\end{split}
\label{lambdaorderf}
\end{equation}
with $\Lambda_{k,l} \geq 0$ a constant parameter for $k,l=1..K$. The `cones' $\mathscr{C}(\Lambda)$ and $\mathscr{C}^{\calligra{o}}(\Lambda)$ are defined in section (3.5). Since $\mathscr{C}(\Lambda)$ and $\mathscr{C}^{\calligra{o}}(\Lambda)$ coincide when $\Lambda=0$, we will assume that $\Lambda_{k,l} > 0$ when $(k,l) \in P$, and leave the case $\Lambda_{k,l}=0$ to $(k,l) \in P^{\calligra{o}}$.

Using the reversed triangular inequality $\sqrt{(a+b)^{2}} \geq \sqrt{a^{2}} + \sqrt{b^{2}}$, one can prove that $\mathscr{C}(\Lambda_{1})+\mathscr{C}(\Lambda_{2}) \subset \mathscr{C}(\Lambda)$ if and only if $\Lambda_{1}+\Lambda_{2} \geq \Lambda$. Using this, one can write the necessary and sufficient conditions on $\preceq_{\Lambda}$ for it to be an order. These are the following:

\begin{itemize}
\item \textit{reflexivity:} $\{(k,k) | k \in \llbracket 1,K \rrbracket\} \subset P^{\calligra{o}}$,   (R)
\item \textit{antisymmetry:} $\forall (k,l), \; k \ne l \Rightarrow (k,l) \in P \text{ or } (l,k) \in P$,   (A)
\item \textit{transitivity:}
\begin{enumerate}
\item $\forall k,l,m: \; \Lambda_{k,l}+\Lambda_{l,m} \geq \Lambda_{k,m}$,   (T1)
\item $\forall (k,l) \in P^{\calligra{o}}, \forall m: \; \Lambda_{l,m} \geq \Lambda_{k,m} \text{ and } \Lambda_{m,k} \geq \Lambda_{m,l}$,   (T2)
\item $\forall k,l,m: \; [(k,l) \in P^{\calligra{o}} \text{ and } (l,m) \in P^{\calligra{o}}] \Rightarrow (k,m) \in P^{\calligra{o}}$.   (T3)
\end{enumerate}
\end{itemize}

Finally, and assuming that $\preceq_{\Lambda}$ is an order, the requirements of Theorem \ref{Class Af 2} are met if and only if:

\begin{equation*}
\forall k: \; n_{k} \geq 2 \Rightarrow \Lambda_{k,k} > 0. \; \text{(O)}
\end{equation*}

Let us give a specific example. Consider the finite algebra $A_{f}=\mathbb{C} \oplus M_{2}(\mathbb{C}) \oplus M_{3}(\mathbb{C})$. First, we have to find a partition $P,P^{\calligra{o}}$ of $\llbracket 1,3 \rrbracket^{2}$ that satisfies rules (R),(A) and (T3). Reflexivity tells us that $(1,1),(2,2),(3,3) \in P^{\calligra{o}}$. We have to put at least half of the remaining couples in $P$ while satisfying antisymmetry. We chose to have $(1,2),(1,3),(3,2) \in P$. What remains has to be split between $P$ and $P^{\calligra{o}}$ in a way that satisfies (T3). For example, we choose to have $(2,3),(3,1) \in P^{\calligra{o}}$, which imposes that the last remaining couple $(2,1)$ be in $P^{\calligra{o}}$ as well. We thus have the partition shown in the left part of Table \ref{Order}.

Next, we need to chose values for the $\Lambda_{k,l}$ that satisfy rules (T1), (T2) and (O). For simplicity, we will choose their values in the set $\{0,\Lambda\}$, where $\Lambda > 0$ is a constant parameter, while keeping as few nonzero $\Lambda$s as possible. We start by filling the diagonal. Because of rule (O), we must choose $\Lambda_{2,2}=\Lambda_{3,3}=\Lambda$. But we are allowed to choose $\Lambda_{1,1}=0$. Using (T2), we see that the inequalities $\Lambda_{k,l} \leq \min(\Lambda_{k,k},\Lambda_{l,l})$ and $\Lambda_{l,k} \geq \max(\Lambda_{k,k},\Lambda_{l,l})$ hold for all $(k,l) \in P^{\calligra{o}}$. For example, we have $(2,1) \in P^{\calligra{o}}$. Thus, we have $\Lambda_{2,1} \leq 0$ and $\Lambda_{1,2} \geq \Lambda$. We choose $\Lambda_{2,1} = 0$ and $\Lambda_{1,2} = \Lambda$. The same can be done for $(3,1)$ and $(2,3)$, leaving us with no more $\Lambda$s to specify. The  chosen $\Lambda$s are displayed in the right part of Table \ref{Order}. We leave it to the reader to check that these values do satisfy rules (T1), (T2) and (O).

\begin{table}[h!]
	\centering
		\begin{tabular}{|c|c|c|c|}
			\hline
 			\backslashbox{k}{l} & 1 & 2 & 3 \\ \hline
 			1 & $P^{\calligra{o}}$ & $P$ & $P$ \\ \hline
 			2 & $P^{\calligra{o}}$ & $P^{\calligra{o}}$ & $P^{\calligra{o}}$ \\ \hline
 			3 & $P^{\calligra{o}}$ & $P$ & $P^{\calligra{o}}$ \\ \hline
		\end{tabular}
		\begin{tabular}{|c|c|c|c|}
			\hline
 			\backslashbox{k}{l} & 1 & 2 & 3 \\ \hline
 			1 & 0 & $\Lambda$ & $\Lambda$ \\ \hline
 			2 & 0 & $\Lambda$ & 0 \\ \hline
 			3 & 0 & $\Lambda$ & $\Lambda$ \\ \hline
		\end{tabular}
	\caption{On the left is displayed the set to which the couple $(k,l)$ belongs. On the right, the value of the parameter $\Lambda_{k,l}$.}
	\label{Order}
\end{table}


\begin{thebibliography}{1234567}

\bibitem [Al-En 05] {ae} S. T. Ali, M. Engli\v{s}, {\it Quantization methods: a guide for physicists and analysts}, Rev. Math. Phys. {\bf 17} (2005) 391 math-ph/0405065
\bibitem [ASZ 14] {asz} N. Alkofer, F. Saueressig, O. Zanusso, \textit{Spectral dimensions from the spectral action}, arXiv: 1410.7999

\bibitem [Ba 07] {barrett} J. Barrett, {\it Lorentzian version of the noncommutative geometry of the standard model of particle physics}, Journal of mathematical physics (2007) hep-th/0608221
\bibitem [Be-Sa 07] {bs} A. N. Bernal, M. {S\'a}nchez, {\it Globally hyperbolic spacetimes can be defined as
`causal' instead of `strongly causal'}, Class. Quantum Grav., {\bf 24} (2007) 745--749 
\bibitem [Be 09] {bes1} F. Besnard, {\it A noncommutative view on topology and order},  J. Geom. Phys. {\bf 59}, pp. 861--875 (2009) abs/0804.3551
\bibitem [Be 12] {bes2} F. Besnard, {\it An approximation theorem for non-decreasing functions on compact posets}, J. Approx. Theory {\bf 172} pp. 1--3 (2013),  abs/1208.2616
\bibitem [Be 13] {bes3} F. Besnard, {\it Noncommutative ordered spaces: examples and counterexamples}, arXiv:1312.2442 
\bibitem [BLMS 87] {sor} L. Bombelli, J. Lee, D. Meyer, and R. D. Sorkin, {\it Space-time as a causal set}, Phys. Rev. Lett. {\bf 59}, 521--524 (1987)
\bibitem [Br 92] {brown} L. Brown, {\it Complements to various Stone-Weierstrass theorems for $C^*$-algebras and a theorem of Shultz}, Commun. Math. Phys. {\bf 143}, 405--413 (1992)
\bibitem [Ch-Co 12] {cc1} A. H. Chamseddine, A. Connes, {\it Resilience of the Spectral Standard Model}, JHEP 1209 (2012) 104
\bibitem [CCS 13] {cc2} A. H. Chamseddine, A. Connes, W. van Suijlekom, {\it Beyond the Spectral Standard Model: Emergence of Pati-Salam Unification}, JHEP 1311 (2013) 132, arXiv:1304.8050
\bibitem [Di 77] {dix} J. Dixmier, {\it $C^*$-algebras}, North-Holland Publishing Company, 1977
\bibitem [DPR 13] {dpr} K. van den Dungen, M. Paschke, A. Rennie, {\it Pseudo-Riemannian spectral triples and the harmonic oscillator}, {J. Geom. Phys.} {\bf 73} (2013), 37-55, arXiv:1207.2112.
\bibitem [Ev-Ok 11] {eo} O. Evren, E.A. Ok, {\it On the  multi-utility representation of preference relations}, Jour. of Math. Econ., {\bf 47}, No 4--5, pp 554--563 (2011)
\bibitem [Fr-Ec 13] {fe} N. Franco and M. Eckstein, {\it Exploring the Causal Structures of Almost Commutative Geometries}, abs/1310.8225 (2013)
\bibitem [KLV 14] {lizzi} M.A. Kurkov, F. Lizzi, D. Vassilevich, {\it High energy bosons do not propagate}, Phys. Lett. B, {\bf 731}, pp 311-315 (2014) 	arXiv:1312.2235
\bibitem [Le 83] {levin} V. L. Levin, {\it A continuous utility theorem for closed pre-order on a $\sigma$-compact metrizable space}, Soviet. Math. Dokl., {\bf 28} (1983) 715--718 
\bibitem [Kr-Pe 67] {kp} E. H. Kronheimer  and R. Penrose, {\it On the structure of causal spaces}, Proc. Camb. Phil. Soc., {\bf 63} (1967) 482--501
\bibitem[Mi 56] {Michael} E. Michael, \textit{Continuous Selections I}, Annals of Mathematics, Second Series, Vol. \textbf{63}, No. 2 (Mar., 1956), pp. 361-382.
\bibitem [Mi-Sa 06] {minguzzisanchez} E. Minguzzi, M. Sànchez, {\it The causal hierarchy of spacetimes} in H. Baum and D. Alekseevsky (eds.), vol. Recent developments in pseudo-Riemannian geometry, ESI Lect. Math. Phys., pp 299 -- 358 (2008) arXiv:gr-qc/0609119
\bibitem [Mi 07] {mingstrK1} E. Minguzzi, {\it The causal ladder and the strength of $K$-causality I}, Class. Quant. Grav. {\bf 25} No 1 (2008)  	arXiv:0708.2070
\bibitem [Mi 09] {minguzzi} E. Minguzzi, {\it Time functions as utilities}, Commun. Math. Phys. {\bf 298} pp 855--868 (2010) arXiv:0909.0890
\bibitem [Pa-Si 06] {ps} M. Paschke , A. Sitarz, {\it Equivariant Lorentzian spectral triples}, math-ph/0611029
\bibitem [Sh-Zo 01] {sz} T. Schucker, S. Zouzou, {\it Spectral action beyond the standard model},  	arXiv:hep-th/0109124
\bibitem [Sh 04] {ncswJ} B. Sheppard, {\it Equivalence of the Stone-Weierstrass conjectures for $C^*$ and $JB^*$-algebras}, Math. Z. {\bf 246}, 105--110 (2004)
\bibitem [St 06] {stro} A. Strohmaier, {\it On noncommutative and pseudo-Riemannian geometry}, { J. Geom. Phys.} {\bf 56} (2006), 175-195. math-ph/0110001
\bibitem [Su 14] {VS} W. van Suijlekom, {\it Noncommutative Geometry and Particle Physics}, Mathematical Physics Studies, Springer 2014.




\end{thebibliography}
\end{document}